\documentclass[
aps,
prc,
showpacs,
twocolumn,
superscriptaddress,
nofootinbib,
floatfix]
{revtex4-1}
\usepackage{graphicx,
longtable,
slashed,
hyperref,
bm,url,
braket,
xcolor,
xspace
}

\usepackage{comment}
\def\be{\begin{equation}}
\def\ee{\end{equation}}

\def\bee{\begin{eqnarray}}
\def\eee{\end{eqnarray}}

\begin{document}

\title{Improved description of the $2\nu\beta\beta$-decay and a possibility to determine
 the effective  axial-vector coupling constant.
}

\author{Fedor \v{S}imkovic}
\affiliation{Department of Nuclear Physics and Biophysics, Comenius
University, Mlynsk\'{a} dolina F1, SK-842 48
Bratislava, Slovakia}
\affiliation{Boboliubov Laboratory of Theoretical Physics, JINR 141980 Dubna,
Russia}
\affiliation{Czech Technical University in Prague, 128-00 Prague, Czech Republic}
\author{Rastislav Dvornick\'{y}}
\affiliation{Department of Nuclear Physics and Biophysics, Comenius
University, Mlynsk\'{a} dolina F1, SK-842 48
Bratislava, Slovakia}
\affiliation{Dzhelepov Laboratory of Nuclear Problems, JINR 141980 Dubna,
Russia}
\author{Du\v{s}an \v Stef\'{a}nik}
\affiliation{Department of Nuclear Physics and Biophysics, Comenius
University, Mlynsk\'{a} dolina F1, SK-842 48
Bratislava, Slovakia}
\author{Amand Faessler}
\affiliation{Institute of Theoretical Physics, University of Tuebingen, Auf der
  Morgenstelle 14, D-72076 Tuebingen, Germany}

\begin{abstract}
  An improved formalism of the two-neutrino double-beta decay
  ($2\nu\beta\beta$-decay) rate is presented, which takes into account the dependence
  of energy denominators on lepton energies via the Taylor expansion. Till now,
  only the leading term in this expansion has been considered.
  The revised $2\nu\beta\beta$-decay rate and differential characteristics depend on
  additional phase-space factors weighted by the ratios of $2\nu\beta\beta$-decay
  nuclear matrix elements with different powers of the energy denominator. For nuclei
  of experimental interest all phase-space factors are calculated by using exact Dirac
  wave functions with finite nuclear size and electron screening. For isotopes with
  measured $2\nu\beta\beta$-decay half-life the involved nuclear matrix elements
  are determined within the quasiparticle random phase approximation with partial
  isospin restoration. The importance of correction terms to the
  $2\nu\beta\beta$-decay rate due to Taylor expansion is established and 
  the modification of shape of single and summed electron energy distributions
  is discussed.  It is found that the improved calculation
  of the $2\nu\beta\beta$-decay predicts slightly suppressed
  $2\nu\beta\beta$-decay background to the neutrinoless double-beta decay signal.
  Further, a novel approach to determine the value of effective weak-coupling constant
  in nuclear medium $g^{\rm eff}_{\rm A}$ is proposed.
\end{abstract}
\medskip

\pacs{
}
\maketitle

%
\section{Introduction}
%

The two-neutrino double beta decay ($2\nu\beta\beta$ decay) \cite{hax84,doi85,ROP12},
\begin{equation}
(A,Z) \rightarrow (A,Z+2) + 2 e^- + 2 {\overline{\nu}}_e,
\end{equation}
a process fully consistent with the standard model of electroweak
interaction, is the rarest process measured so far in the nature. 
It has been observed in twelve even-even nuclei,
in which single-$\beta$ decay is energetically forbidden or strongly
suppressed \cite{barabash}. 

The $2\nu\beta\beta$-decay is a source of background in experiments
looking for a signal of the neutrinoless double beta decay
($0\nu\beta\beta$-decay) \cite{hax84,doi85,ROP12},
\begin{equation}
(A,Z) \rightarrow (A,Z+2) + 2 e^-,
\end{equation}
which observation would prove that neutrinos are
Majorana particles, i.e., their own antiparticles.

The inverse half-life of the $2\nu\beta\beta$ decay is commonly presented
by the product of a phase-space factor $G^{2\nu}$, fourth power of the
effective axial-vector coupling constant
$g^{\rm eff}_A$ and $2\nu\beta\beta$-decay nuclear matrix element (NME)
$M^{2\nu}_{GT}$ as follows:
\begin{equation}
  \left( T^{2\nu}_{1/2}\right)^{-1} = \left(g^{\rm eff}_A\right)^4 ~\left| M^{2\nu}_{GT}\right|^2~ G^{2\nu}.
\end{equation}
The matrix element $M^{2\nu}_{GT}$, which value can be determined from the measured
$2\nu\beta\beta$-half-life by making assumption about the value of
$g_A^{\rm eff}$, plays an important role in understanding of the nuclear structure
of double beta decay isotopes \cite{dussu4}.
Its value is used to adjust the residual part of the nuclear
Hamiltonian in calculation of the $0\nu\beta\beta$-decay NME
within the proton-neutron Quasiparticle Random Phase Approximation (pn-QRPA) \cite{Rod03a,Rod06}. 
Due to this procedure obtained the results are only weakly sensitive on the 
size of the model space and chosen type of NN interaction. So far,
$2\nu\beta\beta$-decay NMEs have been calculated without the closure
approximation only within the Interacting Shell Model (ISM) \cite{LSSM1} and
the pn-QRPA \cite{kmuto}.

The measured single and summed electron differential decay rates
of the $2\nu\beta\beta$-decay allow to get valuable information
concerning many interesting physical issues. In particular, from the shape
of the summed electron distribution we get constraints on the Majoron mode
of the $0\nu\beta\beta$-decay \cite{majoron},
the bosonic neutrino component \cite{bosneu}, violation
of the Lorentz invariance \cite{lorentz}. In addition, a reconstruction of individual
electron energies and angular correlations in the NEMO3 experiment allowed
to obtain information about the Single State Dominance (SSD) and Higher State
Dominance (HSD) hypotheses discussing the importance of various
contributions to the $2\nu\beta\beta$-decay NME from transitions
through intermediate nuclear states \cite{domin1,domin2}. 

Recently, a  significant progress has been achieved in double beta decay experiments.
The $2\nu\beta\beta$-decay mode has been measured with high
statistics in the GERDA ($^{76}$Ge) \cite{expGe}, NEMO3 ($^{100}$Mo) \cite{expMo},
CUORE ($^{130}$Te) \cite{expTe}, 
EXO ($^{136}$Xe)\cite{expXea} and KamlandZEN ($^{136}Xe$) \cite{expXeb} experiments.
As a consequence there is a request for more accurate description of the $2\nu\beta\beta$-decay
process and corresponding differential characteristics. In this contribution
we improve the theoretical description of the $2\nu\beta\beta$-decay process by taking into account
the dependence on lepton energies from the energy denominators of nuclear
matrix elements, which has been neglected till now. In addition, a novel possibility to determine the
effective axial-vector coupling constant $g_A^{\rm eff}$ will be proposed.

%
\section{The improved formalism for description of double-beta decay}
%

In what follows we present improved formulae for the $2\nu\beta\beta$- and $0\nu\beta\beta$-decay
half-lives in which the effect of the lepton energies in the energy denominator of NMEs
is taken into account.

%
\subsection{The $2\nu\beta\beta$-decay rate}
%

The inverse half-life  of the $2\nu\beta\beta$-decay transition to the $0^+$
ground state of the final nucleus takes the form:
\bee
\left[T^{2\nu}_{1/2}\right]^{-1} =
\frac{m_e}{8 \pi^7 \ln{2}}(G_{\beta}m_e^2)^4 \left(g_A^{\rm eff}\right)^4 I^{2\nu},
\eee
where $G_\beta = G_F \cos{\theta_C}$ ($G_F$ is Fermi constant and $\theta_C$ is the Cabbibo angle),
$m_e$ is the mass of electron and
\bee
I^{2\nu} &=&\frac{1}{m_e^{11}}\int_{m_e}^{E_i-E_f-m_e}F_0(Z_f,E_{e_1}) p_{e_1} E_{e_1} dE_{e_1} \nonumber \\
&\times&\int_{m_e}^{E_i-E_f-E_{e_1}} F_0(Z_f,E_{e_2}) p_{e_2} E_{e_2} dE_{e_2} \nonumber\\
&\times&\int_{0}^{E_i-E_f-E_{e_1}-E_{e_2}} E_{\nu_1}^2 E_{\nu_2}^2 {\cal A}^{2\nu} dE_{\nu_1}.
\eee
Here, $E_{\nu_2}=E_i-E_f-E_{e_1}-E_{e_2}-E_{\nu_1}$ due to energy conservation. $E_i$, $E_f$, $E_{e_i}$
($E_{e_i}=\sqrt{p_{e_i}^2+m^2_e}$) and $E_{\nu_i}$ ($i=1,2$) are the energies of initial and final nuclei,
electrons and antineutrinos, respectively.
$F(Z_f, E_{e_i})$ denotes relativistic Fermi function and $Z_f=Z+2$. 
${\cal A}^{2\nu}$ consists of products of the Gamow-Teller nuclear matrix elements (we neglect the
contribution from the double Fermi transitions to the $2\nu\beta\beta$-decay rate),
which depends on  lepton energies \cite{dussu4}:
\bee \label{nucmat}
{\cal A}^{2\nu} &=&  
\left[\frac{1}{4}|M^K_{GT}+M^L_{GT}|^2+\frac{1}{12}|M^K_{GT}-M^L_{GT}|^2\right],
\nonumber
\eee
where
\bee \label{fagtmatt}
M^{K,L}_{GT} &=& m_e \sum_n M_n
\frac{E_n - {(E_i+E_f)}/{2} }{[E_n - {(E_i+E_f)}/{2}]^2 - \varepsilon^2_{K,L}}
\eee
with
\bee
M_n =
\langle 0^+_f \parallel \sum_{m}\tau^-_m \sigma_m \parallel 1^+_{n}\rangle
\langle 1^+_n \parallel \sum_{m}\tau^-_m \sigma_m \parallel 0^+_{i} \rangle, \nonumber\\
\eee
Here, $|0^+_i\rangle$, $|0^+_f\rangle$ are the $0^+$ ground states of the initial and final even-even nuclei,
respectively, and $|1^+_n\rangle$ are all possible states of the intermediate nucleus
with angular momentum and parity $J^\pi = 1^+$ and energy $E_n(1^+)$. The lepton energies
enter in the factors
\begin{eqnarray}
  \varepsilon_{K} &=& \left(E_{e_2}+E_{\nu_2}-E_{e_1}-E_{\nu_1}\right)/2,\nonumber\\
  \varepsilon_{L} &=& \left(E_{e_1}+E_{\nu_2}-E_{e_2}-E_{\nu_1}\right)/2.
\end{eqnarray}  
The maximal value of $|\varepsilon_K|$ and $|\varepsilon_L|$ is the half of $Q$ value of the process
($\varepsilon_{K, L}\in(-Q/2,Q/2)$). For $2\nu\beta\beta$ decay with energetically forbidden transition
to intermediate nucleus ($E_n-E_i > - m_e$) the quantity $E_n -(E_i+E_f)/2 = Q/2 + m_e + (E_n-E_i)$
is always larger than half of the Q value.

The calculation of the $2\nu\beta\beta$-decay probability is usually simplified by an approximation
\bee \label{fagtmat}
&& M^{K,L}_{GT} \simeq M^{2\nu}_{GT} = m_e \sum_n
\frac{M_n}{E_n - (E_i+E_f)/2},\nonumber\\
\eee
which allows a separate calculation of the phase space factor and nuclear matrix element. 

The calculation of $M^{2\nu}_{GT}$ requires to evaluate explicitly the matrix
elements to and from the individual $|1^+_n\rangle$ states
in the intermediate odd-odd nucleus. In the 
IBM calculation of this matrix element \cite{ibmcl} the sum over virtual intermediate nuclear states
is completed by closure after replacing $E_n-(E_i+E_f)/2$ by some average value
$E_{av}$:
\begin{eqnarray}
M^{2\nu}_{GT} &\simeq & \frac{m_e}{E_{av}} {M^{2\nu}_{GT-cl}}
\end{eqnarray}
with
\begin{eqnarray}
M^{2\nu}_{GT-cl}&=& \langle 0^+_f| \sum_{m,n} \tau^-_m \tau^-_n \vec{\sigma}_m\cdot\vec{\sigma}_n| 0^+_i\rangle.
\end{eqnarray}
The validity of the closure approximation is as good as the guess about the average energy
to be used. This approximation might be justified, e.g., in the case  
there is a dominance of transition through a single state of the intermediate nucleus
\cite{abad}.

We get a more accurate expression 
for the $2\nu\beta\beta$-decay rate by performing the
Taylor expansion in matrix elements $M^{K,L}_{GT}$ over the ratio
$\varepsilon_{K,L}/(E_n - {(E_i+E_f)}/{2})$. 
By limiting our consideration to the fourth power in $\varepsilon$ we obtain
\begin{eqnarray}
\left[T^{2\nu}_{1/2}\right]^{-1} \equiv  \frac{\Gamma^{2\nu}}{\ln{(2)}}
\simeq  \frac{\Gamma_0^{2\nu} + \Gamma_2^{2\nu} + \Gamma_4^{2\nu} }{\ln{(2)}},
\label{t2newa}
\end{eqnarray}
where partial contributions to the full $2\nu\beta\beta$-decay width $\Gamma^{2\nu}$ associated with
the leading $\Gamma^{2\nu}_0$, next to leading $\Gamma^{2\nu}_2$ and next-to-next to leading $\Gamma^{2\nu}_4$
orders is Taylor expansion are given by
\begin{eqnarray}
\frac{\Gamma_0^{2\nu}}{\ln{(2)}}
&=& \left(g^{\rm eff}_A\right)^4 {\cal M}_{0} G^{2\nu}_{0}, ~~
\frac{\Gamma_2^{2\nu}}{\ln{(2)}}
= \left(g^{\rm eff}_A\right)^4 {\cal M}_{2} G^{2\nu}_{2}, \nonumber\\
\frac{\Gamma_4^{2\nu}}{\ln{(2)}}
&=& \left(g^{\rm eff}_A\right)^4
\left({\cal M}_{4} G^{2\nu}_{4} + {\cal M}_{22} G^{2\nu}_{22}\right).
\label{t2newb}
\end{eqnarray}
The phase- space factors are defined as 
\begin{eqnarray}
&& G^{2\nu}_N = 
\frac{c_{2\nu}}{m_e^{11}}\int_{m_e}^{E_i-E_f-m_e}F_0(Z_f,E_{e_1}) p_{e_1} E_{e_1} dE_{e_1} \nonumber \\
&&\times\int_{m_e}^{E_i-E_f-E_{e_1}} F_0(Z_f,E_{e_2}) p_{e_2} E_{e_2} dE_{e_2} \\
&&\times\int_{0}^{E_i-E_f-E_{e_1}-E_{e_2}} E_{\nu_1}^2 E_{\nu_2}^2 {\cal A}^{2\nu}_N dE_{\nu_1},~~\textrm{(N=0,~2,~4,~22)}
\nonumber
\eee
with $c_{2\nu} = {m_e} (G_{\beta}m_e^2)^4 /(8 \pi^7 \ln{2})$ and 
\bee
{\cal A}^{2\nu}_0 &=& 1,~~ {\cal A}^{2\nu}_2 = \frac{\varepsilon_K^2 + \varepsilon_L^2}{(2 m_e)^2}, \nonumber\\
{\cal A}^{2\nu}_{22} &=&  \frac{\varepsilon_K^2 \varepsilon_L^2}{(2 m_e)^4},~~
{\cal A}^{2\nu}_4 = \frac{\varepsilon_K^4 + \varepsilon_L^4}{(2 m_e)^4}. \nonumber\\
\eee  
The products of nuclear matrix elements are given by
\begin{eqnarray}
&&{\cal M}_{0} = \left( M^{2\nu}_{GT-1}\right)^2, \nonumber\\
&&{\cal M}_{2} = M^{2\nu}_{GT-1} M^{2\nu}_{GT-3}, \nonumber\\
&&{\cal M}_{22} =  \frac{1}{3} \left( M^{2\nu}_{GT-3} \right)^2, \nonumber\\
&&{\cal M}_{4} =  \frac{1}{3} \left( M^{2\nu}_{GT-3}\right)^2
+ M^{2\nu}_{GT-1} M^{2\nu}_{GT-5},
\end{eqnarray}
where nuclear matrix elements take the forms
\begin{eqnarray}
  && M^{2\nu}_{GT-1} \equiv M^{2\nu}_{GT} \nonumber\\
  && M^{2\nu}_{GT-3} = \sum_n M_n \frac{4~ m_e^3}{\left(E_n - (E_i+E_f)/2\right)^3},\nonumber\\
  && M^{2\nu}_{GT-5} = \sum_n M_n \frac{16~ m_e^5}{\left(E_n - (E_i+E_f)/2\right)^5}.
  \label{2nnme}
\end{eqnarray}

By introducing two ratios of nuclear matrix elements,
\begin{eqnarray}
  \xi^{2\nu}_{31}  = \frac{M^{2\nu}_{GT-3}}{M^{2\nu}_{GT-1}},~~
  \xi^{2\nu}_{51} = \frac{M^{2\nu}_{GT-5}}{M^{2\nu}_{GT-1}},
\label{ratioxi}
\end{eqnarray}  
the $2\nu\beta\beta$-decay half-life,
\begin{eqnarray}
  &&  \left[T^{2\nu \beta \beta}_{1/2}\right]^{-1} = \left(g^{\rm eff}_A\right)^4 \left| M^{2\nu}_{GT-1}\right|^2
  \left(G^{2\nu}_{0}  + \xi^{2\nu}_{31}  G^{2\nu}_{2}\right.
  \nonumber\\
  && ~~\left.
  + \frac{1}{3} \left(\xi^{2\nu}_{31}\right)^2 G^{2\nu}_{22} +
  \left(\frac{1}{3} \left(\xi^{2\nu}_{31}\right)^2 + \xi^{2\nu}_{51}\right) G^{2\nu}_{4}\right),
\end{eqnarray}
is expressed with single NME ($M^{2\nu}_{GT-1}$) and two ratios of nuclear 
matrix elements  ($\xi^{2\nu}_{31}$ and $\xi^{2\nu}_{51}$),
which have to be calculated by means of the nuclear structure theory,
four phase-space factors ($G^{2\nu}_0$, $G^{2\nu}_2$, $G^{2\nu}_{22}$
and $G^{2\nu}_4$), which can be computed with a good accuracy,  and the
unknown parameter $g_A^{\rm eff}$.

%
\subsection{$0\nu\beta\beta$-decay rate}
%

The inverse lifetime of the $0\nu\beta\beta$ decay is commonly presented as
a product of the total lepton number violating Majorana neutrino mass
$m_{\beta\beta}$, the phase-space factor $G^{0\nu}$, nuclear matrix element  
${M'}^{0\nu}(g_A^{\rm eff})$ and unquenched axial-vector
coupling constant $g_A$ ($g_A$ = 1.269) in fourth power as follows \cite{ROP12}:
\begin{equation}
  \left({T^{0\nu}_{1/2}}\right)^{-1} = \left|\frac{m_{\beta\beta}}{m_e}\right|^2~g^4_A
  ~\left|{M'}^{0\nu}(g_A^{\rm eff})\right|^2~G^{0\nu},
\label{0nhalflife}
\end{equation}
where 
\bee 
&& G^{0\nu} = \frac{G_{\beta}^{4}m_e^7}{32  \pi^5 R^2 \ln{(2)}}\frac{1}{m_e^5}\times \\ 
&& \int_{m_e}^{E_i-E_f-m_e}~
F_0(Z_f,E_{e_1}) p_{e_1} E_{e_1}
F_0(Z_f,E_{e_2}) p_{e_2} E_{e_2} dE_{e_1} \nonumber
\eee
with $E_{e_2} = E_i-E_f-E_{e_1}$, $p_{e_i}=\sqrt{E_{e_i}^2-m_e^2}$ (i=1,2). 
The NME takes the form 
\begin{eqnarray}
\label{eq:0nume}
&&      {M'}^{0\nu} (g_A^{\rm eff}) = \frac{R}{2 \pi^2 g_A^2} \times \nonumber\\
&&\sum_{n} 
\int e^{i\mathbf{p}\cdot (\mathbf{x}-\mathbf{y})}
\frac{
\langle 0^+_f| {J}^{\mu\dag}_L(\mathbf{x})|n\rangle
\langle n|{J}^\dag_{L \mu} (\mathbf{y}) |0^+_i\rangle}
  {p (p+E_n-\frac{E_i-E_f}{2})}
  d^3p~ d^3x~ d^3y.
\label{0nbbNMEa}\nonumber \\
\end{eqnarray}
We note that the axial-vector $g_A^{\rm eff}(p^2)$ and induced pseudoscalar $g_P^{\rm eff}(p^2)$ 
form factors of nuclear hadron currents ${J}^{\mu\dag}$ are
``renormalized in nuclear medium''. The magnitude and origin
of this renormalization is the subject of the  analysis of many works, since it
tends to increase the $0\nu\beta\beta$-decay half-life in comparison with the case in which this
effect is absent \cite{javierga,engelga}. 

In derivation of the $0\nu\beta\beta$-decay rate in Eq. (\ref{0nhalflife})
the standard approximations were adopted: i) a factorization of phase-space factor
and nuclear matrix element was achieved by approximation,
in which electron wave functions were
replaced by their values at the nuclear radius R.
ii) the  dependence on lepton energies in energy denominators
of the $0\nu\beta\beta$-decay NME was neglected.

Here, we go beyond the approximation ii). The $0\nu\beta\beta$ nuclear matrix element 
contains a sum of two energy denominators:
\begin{equation}
\frac{1}{p_0 + E_n - E_i + E_{e_1}} + \frac{1}{p_0 + E_n - E_i + E_{e_2}},
\end{equation}
where $p = (p_0, \, \bm{p})$ is the four-momentum transferred by the Majorana neutrino
(common for all neutrino mass eigenstates, since the neutrino masses $m_i$ can be safely neglected
in $p_0 = \sqrt{\vec{p}^2 + m_i^2} \approx |\vec{p}| \sim 100 \, \mathrm{MeV}$).
By taking advantage of the energy conservation $E_i = E_f + E_{e_1} + E_{e_2}$ (the effect of nuclear
recoil is disregarded) the approximation was adopted as follows:
\begin{equation}
  \frac{2 \left( p_0 + E_n - \frac{E_i + E_f}{2} \right)}
       {\left( p_0 + E_n - \frac{E_i + E_f}{2} \right)^2 -  \varepsilon^2}
       \simeq \frac{2}{p_0 + E_n - \frac{E_i + E_f}{2}}
\label{app0n}
\end{equation}
with $\varepsilon = (E_{e_1}-E_{e_2})/2$. 
More accurate expression for the $0\nu\beta\beta$-decay half-life is achieved
by taking into account next term in Taylor expansion over the quantity 
$\varepsilon^2/[p_0 + E_n - (E_i + E_f)/2]^2$ in Eq. (\ref{app0n}). We end up with 
\begin{eqnarray}
\left({T^{0\nu}_{1/2}}\right)^{-1} = \left|\frac{m_{\beta\beta}}{m_e}\right|^2~g^4_A 
\left|{M'}^{0\nu}_1 \right|^2
\left(G^{0\nu}_0~ +~ 2~\xi^{0\nu}_{31}~ G^{0\nu}_2\right),\nonumber\\
\label{hl0neps}
\end{eqnarray}
where
\bee 
&& G^{0\nu}_N = \frac{G_{\beta}^{4}m_e^7}{32  \pi^5 R^2 \ln{(2)}}\frac{1}{m_e^5}\times \\ 
&& \int_{m_e}^{E_i-E_f-m_e}~{\cal A}^{0\nu}_N
F_0(Z_f,E_{e_1}) p_{e_1} E_{e_1}
F_0(Z_f,E_{e_2}) p_{e_2} E_{e_2} dE_{e_1} \nonumber
\eee
with
\begin{equation}
{\cal A}^{0\nu}_0 = 1, ~~~~~~~~{\cal A}^{0\nu}_2 = \varepsilon^2/(2 m_e)^2.
\end{equation}  
The additional term in the $0\nu\beta\beta$-decay rate in Eq. (\ref{hl0neps}) 
is weighted by ratio $\xi^{0\nu}_{31}$,
\begin{equation}
  \xi^{0\nu}_{31} = \frac{{M'}^{0\nu}_3(g_A^{\rm eff})}{{M'}^{0\nu}_1 (g_A^{\rm eff})}.
  \label{0nratio}
\end{equation}  
of two NMEs defined as follows:
\begin{eqnarray}
  && {M'}^{0\nu}_1 (g_A^{\rm eff}) \equiv {M'}^{0\nu} (g_A^{\rm eff}) \nonumber\\
  &&      {M'}^{0\nu}_3 (g_A^{\rm eff})
  = \frac{R}{2 \pi^2 g_A^2} (2 m_e)^2\times \nonumber\\
&& \sum_{n} 
\int e^{i\mathbf{p}\cdot (\mathbf{x}-\mathbf{y})}
\frac{
\langle 0^+_f| {J}^{\mu\dag}_L(\mathbf{x})|n\rangle
\langle n|{J}^\dag_{L \mu} (\mathbf{y}) |0^+_i\rangle}
  {p \left( p + E_n - \frac{E_i-E_f}{2} \right)^3}
  d^3p~ d^3x~ d^3y.
  \label{0nbbNMEb}
  \nonumber \\
\end{eqnarray}

\begin{widetext}
  
\begin{table}[!t]
  \begin{center}
    \caption{Phase-space factors $G^{2\nu}_{0, 2, 22, 4}$ ($G^{0\nu}_{0, 2}$)
      entering the $2\nu\beta\beta$-decay ($0\nu\beta\beta$-decay) rate in
      Eq. (\ref{t2newa}) (Eq. (\ref{hl0neps}). The radial wave-functions
      $g_{-1}$ and $f_{+1}$ of an electron, which constitute the Fermi function
      in Eq. (\ref{fermif}), were calculated in two approximation
      schemes: (A) The standard approximation of Doi et al. \cite{doi85}.
      (B) The exact solution for a Dirac equation for a uniform charge distribution
      in nucleus and electron screening is taken into account.
\label{tab:phas}}
\centering 
\renewcommand{\arraystretch}{1.1}  
      \begin{tabular}{lcccccccc}\hline\hline
        &          &  \multicolumn{4}{c}{ $2\nu\beta\beta$-decay } & & \multicolumn{2}{c}{ $0\nu\beta\beta$-decay }
     \\ \cline{3-6} \cline{8-9}
nucl. & el. w.f. & $G^{2\nu}_{0}$ [yr$^{-1}$] & $G^{2\nu}_{2}$ [yr$^{-1}$] & $G^{2\nu}_{4}$ [yr$^{-1}$] & $G^{2\nu}_{22}$ [yr$^{-1}$] & &
        $G^{0\nu}_{0}$ [yr$^{-1}$] & $G^{0\nu}_{2}$ [yr$^{-1}$] \\ \hline
 $^{48}$Ca & A &  $1.608\times 10^{-17}$ &  $1.372\times 10^{-17}$  & $1.484\times 10^{-17}$  & $3.297\times 10^{-18}$ & &  $2.641\times 10^{-14}$ &  $2.284\times 10^{-14}$  \\
           & B &  $1.534\times 10^{-17}$ &  $1.307\times 10^{-17}$  & $7.064\times 10^{-18}$  & $3.140\times 10^{-18}$ & &  $2.489\times 10^{-14}$ &  $2.150\times 10^{-14}$  \\
 $^{76}$Ge & A &  $5.278\times 10^{-20}$ &  $1.113\times 10^{-20}$  & $2.924\times 10^{-21}$  & $6.898\times 10^{-22}$ & &  $2.613\times 10^{-15}$ &  $6.269\times 10^{-16}$  \\
           & B &  $4.816\times 10^{-20}$ &  $1.015\times 10^{-20}$  & $1.332\times 10^{-21}$  & $6.284\times 10^{-22}$ & &  $2.370\times 10^{-15}$ &  $5.670\times 10^{-16}$  \\
 $^{82}$Se & A &  $1.763\times 10^{-18}$ &  $7.805\times 10^{-19}$  & $4.333\times 10^{-19}$  & $9.912\times 10^{-20}$ & &  $1.147\times 10^{-14}$ &  $5.449\times 10^{-15}$  \\
           & B &  $1.591\times 10^{-18}$ &  $7.037\times 10^{-19}$  & $1.952\times 10^{-19}$  & $8.931\times 10^{-20}$ & &  $1.020\times 10^{-14}$ &  $4.830\times 10^{-15}$  \\
 $^{96}$Zr & A &  $7.777\times 10^{-18}$ &  $4.292\times 10^{-18}$  & $2.974\times 10^{-18}$  & $6.774\times 10^{-19}$ & &  $2.423\times 10^{-14}$ &  $1.422\times 10^{-14}$  \\
           & B &  $6.796\times 10^{-18}$ &  $3.745\times 10^{-18}$  & $1.296\times 10^{-18}$  & $5.907\times 10^{-19}$ & &  $2.067\times 10^{-14}$ &  $1.209\times 10^{-14}$  \\
$^{100}$Mo & A &  $3.818\times 10^{-18}$ &  $1.747\times 10^{-18}$  & $1.001\times 10^{-18}$  & $2.301\times 10^{-19}$ & &  $1.890\times 10^{-14}$ &  $9.357\times 10^{-15}$  \\
           & B &  $3.303\times 10^{-18}$ &  $1.509\times 10^{-18}$  & $4.320\times 10^{-19}$  & $1.986\times 10^{-19}$ & &  $1.599\times 10^{-14}$ &  $7.886\times 10^{-15}$  \\  
$^{110}$Pd & A &  $1.629\times 10^{-19}$ &  $3.405\times 10^{-20}$  & $8.832\times 10^{-21}$  & $2.115\times 10^{-21}$ & &  $5.783\times 10^{-15}$ &  $1.408\times 10^{-15}$  \\
           & B &  $1.379\times 10^{-19}$ &  $2.881\times 10^{-20}$  & $3.735\times 10^{-21}$  & $1.789\times 10^{-21}$ & &  $4.833\times 10^{-15}$ &  $1.172\times 10^{-15}$  \\
$^{116}$Cd & A &  $3.314\times 10^{-18}$ &  $1.318\times 10^{-18}$  & $6.546\times 10^{-19}$  & $1.522\times 10^{-19}$ & &  $2.064\times 10^{-14}$ &  $9.061\times 10^{-15}$  \\
           & B &  $2.763\times 10^{-18}$ &  $1.097\times 10^{-18}$  & $2.722\times 10^{-19}$  & $1.266\times 10^{-19}$ & &  $1.677\times 10^{-14}$ &  $7.334\times 10^{-15}$  \\
$^{124}$Sn & A &  $6.717\times 10^{-19}$ &  $1.794\times 10^{-19}$  & $5.954\times 10^{-20}$  & $1.414\times 10^{-20}$ & &  $1.124\times 10^{-14}$ &  $3.442\times 10^{-15}$  \\ 
           & B &  $5.534\times 10^{-19}$ &  $1.476\times 10^{-19}$  & $2.448\times 10^{-20}$  & $1.163\times 10^{-20}$ & &  $9.077\times 10^{-15}$ &  $2.768\times 10^{-15}$  \\
$^{128}$Te & A &  $3.314\times 10^{-22}$ &  $1.314\times 10^{-23}$  & $6.409\times 10^{-25}$  & $1.688\times 10^{-25}$ & &  $7.263\times 10^{-16}$ &  $3.875\times 10^{-17}$  \\
           & B &  $2.699\times 10^{-22}$ &  $1.070\times 10^{-23}$  & $2.609\times 10^{-25}$  & $1.374\times 10^{-25}$ & &  $5.904\times 10^{-16}$ &  $3.145\times 10^{-17}$  \\
$^{130}$Te & A &  $1.885\times 10^{-18}$ &  $6.112\times 10^{-19}$  & $2.467\times 10^{-19}$  & $5.812\times 10^{-20}$ & &  $1.807\times 10^{-14}$ &  $6.619\times 10^{-15}$  \\
           & B &  $1.530\times 10^{-18}$ &  $4.953\times 10^{-19}$  & $9.985\times 10^{-20}$  & $4.707\times 10^{-20}$ & &  $1.428\times 10^{-14}$ &  $5.212\times 10^{-15}$  \\
$^{134}$Xe & A &  $2.924\times 10^{-22}$ &  $1.066\times 10^{-23}$  & $4.773\times 10^{-25}$  & $1.264\times 10^{-25}$ & &  $7.613\times 10^{-16}$ &  $3.761\times 10^{-17}$  \\  
           & B &  $2.347\times 10^{-22}$ &  $8.553\times 10^{-24}$  & $1.915\times 10^{-25}$  & $1.014\times 10^{-25}$ & &  $6.100\times 10^{-16}$ &  $3.008\times 10^{-17}$  \\
$^{136}$Xe & A &  $1.793\times 10^{-18}$ &  $5.516\times 10^{-19}$  & $2.110\times 10^{-19}$  & $4.994\times 10^{-20}$ & &  $1.881\times 10^{-14}$ &  $6.590\times 10^{-15}$  \\
           & B &  $1.433\times 10^{-18}$ &  $4.404\times 10^{-19}$  & $8.417\times 10^{-20}$  & $3.986\times 10^{-20}$ & &  $1.464\times 10^{-14}$ &  $5.107\times 10^{-15}$  \\
$^{150}$Nd & A &  $4.817\times 10^{-17}$ &  $2.731\times 10^{-17}$  & $1.937\times 10^{-17}$  & $4.479\times 10^{-18}$ & &  $8.827\times 10^{-14}$ &  $5.462\times 10^{-14}$  \\
           & B &  $3.642\times 10^{-17}$ &  $2.061\times 10^{-17}$  & $7.295\times 10^{-18}$  & $3.380\times 10^{-18}$ & &  $6.339\times 10^{-14}$ &  $3.903\times 10^{-14}$  \\
\hline\hline
      \end{tabular}
  \end{center}
\end{table}     

\end{widetext}      

\begin{widetext}

\begin{table}[!t]
  \begin{center}
    \caption{The $2\nu\beta\beta$- and $0\nu\beta\beta$-decay nuclear matrix elements and ratios
      of nuclear matrix elements (see Eq. (\ref{ratioxi}) and Eq. (\ref{0nratio}))
      calculated within the pn-QRPA with partial isospin restoration \cite{vadimp}.
      $P^{2\nu}_0$, $P^{2\nu}_2$ and $P^{2\nu}_4$ are the leading first, second and third order term contributions
      to the $2\nu\beta\beta$-decay rate in the Taylor expansion. $T^{2\nu-exp}_{1/2}$  is the
      averaged value of the $2\nu\beta\beta$-decay half-life \cite{barabash} considered in
      the calculation of the $2\nu\beta\beta$-decay NMEs. $g^{\rm eff}_A$
      is the effective axial-vector coupling constant.
\label{tab:nme}}
\centering 
\renewcommand{\arraystretch}{1.1}  
      \begin{tabular}{lccccccccccccc}\hline\hline
        &          &  \multicolumn{9}{c}{ $2\nu\beta\beta$-decay } & & \multicolumn{2}{c}{ $0\nu\beta\beta$-decay }
        \\ \cline{3-11} \cline{13-14}
        nucl. & $g^{\rm eff}_A$ & $M^{2\nu}_{GT-1}$ & $M^{2\nu}_{GT-3}$ & $M^{2\nu}_{GT-5}$ &
        $\xi^{2\nu}_{31}$ & $\xi^{2\nu}_{51}$ &
        $P_0^{2\nu}$ & $P_2^{2\nu}$ & $P_4^{2\nu}$ & $T^{2\nu-exp}_{1/2}$ [yr] &  & 
        ${M'}^{0\nu}_1$ & $\xi^{0\nu}_{31}$ \\ \hline
 $^{48}$Ca
        &  0.800  & 0.0553 & 0.0105 & 0.00163  &  0.1891 & 0.0295 & 0.8456 & 0.1362 & 0.0182 & $4.4\times 10^{19}$
        & & 0.4066 & $6.463\times 10^{-4}$    \\
        &  1.000  & 0.0352 & 0.00723 & 0.00105 & 0.2055 & 0.0298 & 0.8346 & 0.1461 & 0.0193 &
        & & 0.4543 & $6.732\times 10^{-4}$     \\
        &  1.269  & 0.0214 & 0.00539 & 0.00075 & 0.2514 & 0.0351 & 0.8036 & 0.1722 & 0.0242 &
        & & 0.5288 &	$6.814\times 10^{-4}$      \\
 $^{76}$Ge
        &  0.800  & 0.175 & 0.0214   & 0.00445 & 0.1220 & 0.0254 & 0.9741  & 0.0250 & 0.0009 & $1.65\times 10^{21}$
        & & 3.1822 & $2.629\times 10^{-4}$   \\
        &  1.000  & 0.111 & 0.0133   & 0.00263 &  0.1204 & 0.0237 & 0.9745  & 0.0247 & 0.0008 &
        & &  3.8830 & $2.484\times 10^{-4}$    \\
        &  1.269  & 0.689 & 0.00716  & 0.00716 &  0.1040 & 0.0170 & 0.9780  & 0.0214 & 0.0006 &
        & &  5.1527 & $2.282\times 10^{-4}$    \\
 $^{82}$Se
        &  0.800  & 0.124  & 0.0216  & 0.00645  & 0.1745 & 0.0521 & 0.9213  & 0.0711 & 0.0076 & $0.92\times 10^{20}$
        & & 2.7859 & $2.243\times 10^{-4}$ 	\\
        &  1.000  & 0.0795 & 0.0129  &  0.00355 & 0.1620 & 0.0446 & 0.9271  & 0.0664 & 0.0065 &
        & & 3.4668 & $2.146\times 10^{-4}$      \\
        &  1.269  & 0.0498 & 0.00643  & 0.00136 & 0.1290 & 0.0272 & 0.9421  & 0.0538 & 0.0041 &
        & & 4.6511 & $2.020\times 10^{-4}$      \\
 $^{96}$Zr
        &  0.800  & 0.1146 & 0.0348   & 0.00885 & 0.3036 & 0.0773 & 0.8399  & 0.1405 & 0.0195 & $2.3\times 10^{19}$
        & & 1.9299 & $6.872\times 10^{-4}$  	\\
        &  1.000  & 0.0718 & 0.273    & 0.00697 & 0.3800 & 0.0971 & 0.8056  & 0.1687 & 0.0257 &
        & & 2.2449 & $8.552\times 10^{-4}$      \\
        &  1.269  & 0.0431 & 0.0220   & 0.00564 & 0.5101 & 0.1309 & 0.7518  & 0.2113 & 0.0369 &
        & & 2.8163 & $1.009\times 10^{-3}$      \\
 $^{100}$Mo
        &  0.800  & 0.292  & 0.123    &  0.0453 & 0.4230 & 0.1553  & 0.8163 & 0.1578 & 0.0259 & $7.1\times 10^{18}$
        & & 3.4765 & $8.297\times 10^{-4}$  	\\
        &  1.000  & 0.184  & 0.0876   & 0.0322 & 0.4752 & 0.1745  & 0.7972 & 0.1731 & 0.0297 &
        & & 4.1737 & $8.997\times 10^{-4}$      \\
        &  1.269  & 0.112  &  0.0633  & 0.0233 & 0.5646 & 0.2075  & 0.7661 & 0.1976 & 0.0363 &
        & & 5.3824 & $8.908\times 10^{-4}$      \\
 $^{116}$Cd
        &  0.800  & 0.1653 & 0.0478   & 0.0142  & 0.2890 & 0.0857  & 0.8872  & 0.1018 & 0.0110  & $2.87\times 10^{19}$
        & & 2.5488 & $4.930\times 10^{-4}$  	\\
        &  1.000  & 0.1053 & 0.0327   &  0.00972 & 0.3102 & 0.0923  & 0.8796  & 0.1083 & 0.0121  &
        & &  3.0859 & $5.240\times 10^{-4}$     \\
        &  1.269  & 0.0651 & 0.0219   & 0.00654 & 0.3370 & 0.1000  & 0.8702  & 0.1164 & 0.0134  &
        & &  4.0381 & $4.998\times 10^{-4}$     \\
 $^{130}$Te
        &  0.800  & 0.0466 & 0.00873  & 0.00239 & 0.1873 & 0.0512 & 0.9389 & 0.0569  & 0.0042  & $6.9\times 10^{20}$
        & & 2.4122 & $4.830\times 10^{-4}$  	\\
        &  1.000  & 0.0298 & 0.00577  & 0.00144 & 0.1937 & 0.0482 & 0.9371 & 0.0588  & 0.0041  &
        & & 2.9617 & $2.629\times 10^{-4}$      \\
        &  1.269  & 0.0185 & 0.00373  &  0.00078 & 0.2015 & 0.0420 & 0.9352 & 0.0610  & 0.0038  &
        & & 3.9026 & $1.488\times 10^{-4}$      \\
 $^{136}$Xe
        &  0.800  & 0.0268 & 0.00706  & 0.00232  & 0.2637 & 0.0866 & 0.9190 & 0.0745 & 0.0065 & $2.19\times 10^{21}$
        & & 1.3425 & $1.608\times 10^{-4}$  	\\
        &  1.000  & 0.0170 & 0.00526 & 0.00169 & 0.3098 & 0.0995 & 0.9059 & 0.0863 & 0.0078 &
        & & 1.6525 & $1.561\times 10^{-4}$      \\
        &  1.269  & 0.0104 & 0.00403 & 0.00126  & 0.3867 & 0.1207 & 0.8848 & 0.1051 & 0.0101 &
        & & 2.1841 & $1.509\times 10^{-4}$      \\
 \hline\hline
      \end{tabular}
  \end{center}
\end{table}     

\end{widetext}

%
\section{Calculations and results}
%

\subsection{Phase-space factors and the QRPA NMEs}

The $2\nu\beta\beta$- and $0\nu\beta\beta$-phase-space factors
presented in the previous section are associated with the
$s_{1/2}$ electron wave function distorted by the Coulomb field:
\begin{eqnarray}
\Psi^{(s_{1/2})}(E_e,\bm{r}) &=& 
\left(\begin{array}{l}
  g_{-1 }(E_e,r)  \chi_s \\
f_{+1 }(E_e,r) (\mathbf{\sigma\cdot\hat{p}_e}) \chi_s
\end{array}\right), 
\end{eqnarray}
where $E_e$ and $\bm{p}_e$ are  the electron energy and momentum, respectively.
$\hat{p}_e=\bm{p}_e/|\bm{p}_e|$ and $r=|\bm{r}|$ is the radial coordinate of the
position of the electron. The values of the radial functions
$g_{-1 }(E_e,r)$ and $f_{+1 }(E_e,r)$  at nuclear radius $r=R$ constitute
the Fermi function as follows:
\begin{equation}
F_0(Z,E_e) = g^2_{-1 }(E_e,R) + f^2_{+1 }(E_e,R). 
\label{fermif}
\end{equation}

Two different approximation schemes for the calculation of radial wave functions
$g_{-1}(E_e,R)$ and $f_{+1}(E_e,R)$ are considered.\\

{\it The  approximation scheme A):} The relativistic electron wave function in a uniform 
charge distribution in nucleus is considered. The lowest terms in the power expansion
in r/R are taken into account. The Fermi function takes the form
\begin{eqnarray}
F_{0} &=& \left[ \frac{\Gamma(3)}{\Gamma(1)\Gamma(1+2\gamma_{0})}
\right]^2 (2 p_e R)^{2(\gamma_{0}-1)} e^{\pi y} 
\mid \Gamma(\gamma_{0}+iy)  \mid^2,\nonumber\\
\end{eqnarray}
where $\gamma_{0} = \sqrt{1-(\alpha )^2}$  and $y = \alpha Z \frac{\varepsilon}{p_e}$.

{\it The  approximation scheme B):} The exact Dirac wave functions
with finite nuclear size and electron screening are used \cite{iachello}. 
The effect  of screening of atomic electrons is taken into account
by the Thomas-Fermi approximation. The numerical calculation is
accomplished by the subroutine package RADIAL \cite{radial}.  

In Table \ref{tab:phas} the $2\nu\beta\beta$- and $0\nu\beta\beta$-decay phase-space factors
calculated within approximations A and B are presented for 13 isotopes of experimental interest.
We see that all phase-space factors calculated with exact relativistic electron wave functions
(the approximation scheme B) are smaller in comparison with those obtained in approximation
scheme A. We note that in both approximation schemes the factorization of phase-space factors
and nuclear matrix elements is achieved by considering radial electron wave functions
at nuclear radius and the difference between them is due to a different treatment of the
Coulomb interaction.

In what follows entries B from Table \ref{tab:phas} will be used in calculation of the
$2\nu\beta\beta$ differential characteristics and decay rates. 

\begin{figure}[!t]
  \includegraphics[width=1.05\columnwidth]{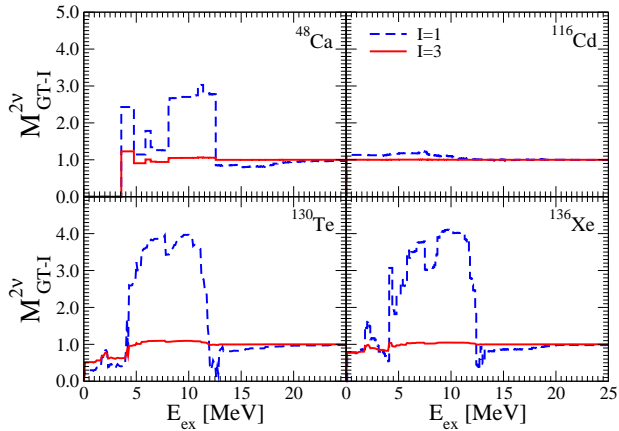}
  \caption{(Color online) Running sum of the $2\nu\beta\beta$-decay NMEs
    $M^{2\nu}_{GT-1}$ and $M^{2\nu}_{GT-3}$ (see Eq. (\ref{2nnme})) for
    $^{48}$Ca, $^{116}$Cd, $^{130}$Te and $^{136}$Xe  (normalized to unity)
    as a function of the excitation energy $E_{ex}$ counted from the
    ground state of intermediate nucleus.  Calculations were performed
    within the proton-neutron QRPA with isospin restoration \cite{vadimp}. Results are
    obtained with Argonne V18 potential and for unquenched axial vector coupling constant
    $g_A=1.269$.
  }
\label{fig.runb}
\end{figure}

\subsection{Nuclear matrix elements}

The $2\nu\beta\beta$- and $0\nu\beta\beta$-decay nuclear matrix elements
(see Eqs. (\ref{2nnme}) and and (\ref{0nbbNMEb}))
are calculated within the proton-neutron quasiparticle random phase approximation
(QRPA) with isospin restoration \cite{vadimp}. 
They were obtained by considering the same model spaces and mean fields as in \cite{vadimp}. The
G-matrix elements of a realistic Argonne V18 nucleon-nucleon potential are considered.
By using the improved theoretical description of the $2\nu\beta\beta$-decay
rate in Eqs. (\ref{t2newa})-(\ref{2nnme}) the isoscalar neutron-proton interaction of
the nuclear Hamiltonian is adjusted to reproduce correctly the average $2\nu\beta\beta$-decay
half-live \cite{barabash} for each nucleus and each $g^{\rm eff}_A$.

In Table \ref{tab:nme} calculated $2\nu\beta\beta$-deacy NMEs are presented
for $g^{\rm eff}_A$= 0.8, 1.0 and 1.269 (unquenched value).
We see that for all isotopes the inequality $M^{2\nu}_{GT-1} > M^{2\nu}_{GT-3} >  M^{2\nu}_{GT-5}$
is valid. The ratios of nuclear matrix elements $\xi^{2\nu}_{31}$, $\xi^{2\nu}_{51}$ and $\xi^{0\nu}$
depend only weakly on $g^{\rm eff}_A$. The largest values $\xi^{2\nu}_{31}=$ 0.56 and
$\xi^{2\nu}_{51}$ = 0.21 are in the case  $^{100}$Mo.

The ratio $\xi^{2\nu}_{31}$ of nuclear matrix element $M^{2\nu}_{GT-3}$ and $M^{2\nu}_{GT-1}$ 
(see Eq. (\ref{ratioxi})) is an important quantity due to a different
structure of both nuclear matrix elements. This fact is
displayed in Fig. \ref{fig.runb} (Fig. \ref{fig.runa}), where
running sum of matrix elements $M^{2\nu}_{GT-1}$
and  $M^{2\nu}_{GT-3}$ is plotted as a function of the excitation energy
$E_{ex}$ counted from the ground state of the intermediate nucleus 
for the $2\nu\beta\beta$-decay of $^{76}$Ge, $^{82}$Se, $^{96}$Zr and $^{100}$Mo
($^{48}$Ca, $^{116}$Cd, $^{130}$Te and $^{136}$Xe).  The results were 
obtained within the QRPA with partial isospin restoration \cite{vadimp}.
By glancing these figures we see that matrix element  $M^{2\nu}_{GT-3}$
is determined by transitions through the lightest states of the intermediate nucleus
unlike $M^{2\nu}_{GT-1}$, which depends also on the transitions through
higher lying states even from the region of Gamow-Teller resonance and
a mutual cancellation among different contributions.

The convergence of the Taylor expansion of the $2\nu\beta\beta$-decay rate
(see Eqs. (\ref{t2newa})-(\ref{2nnme}) depends on values of original
$M^{2\nu}_{GT-1}$ and new $M^{2\nu}_{GT-3,5}$ nuclear matrix elements.
Recall that the powers of $\varepsilon_{K,L}$ are included in
the generalized phase space factors $G^{2\nu}_{0,2,22,4}$ and the denominators are
included in the new nuclear matrix elements $M^{2\nu}_{GT-3,5}$. The leading first $P^{2\nu}_0$,
second $P^{2\nu}_2$ and third $P^{2\nu}_4$ order term contributions to the $2\nu\beta\beta$-decay rate
in the Taylor expansion normalized to the full decay rate are defined as
\begin{eqnarray}
P^{2\nu}_I = \frac{\Gamma^{2\nu}_I}{\Gamma^{2\nu}} 
\end{eqnarray}
with I=0, 2 and 4. Their values calculated with help of the $2\nu\beta\beta$-decay
NMEs evaluated within the QRPA with partial restoration of isospin symmetry \cite{vadimp}
are shown in Table \ref{tab:nme}. We notice a good convergence of contributions to the
$2\nu\beta\beta$-decay rate  due to the Taylor expansion.
The size of these corrections depends on a given isotope.  The largest value of about
25\% is found by $^{100}$Mo. 

In the Table \ref{tab:nme} the calculated $0\nu\beta\beta$-decay nuclear matrix elements
are  presented as well. They were obtained under common assumption that 
the same $g_A^{\rm eff}$ governs both modes of double beta decay \cite{ROP12,vadimp}.
The modification of the $2\nu\beta\beta$-decay rate due to the Taylor expansion has only 
negligible effect on calculation of the $0\nu\beta\beta$-decay NMEs ${M'}^{0\nu}_{1, 3}$
in the context of adjusting the particle-particle interaction strength.  

By glancing the Table \ref{tab:nme} we see that the value of $\xi^{0\nu}_{31}$ is
very small, namely significantly smaller as $\xi^{2\nu}_{31}$,
as the average momentum of neutrino entering the energy denominator
in Eq. (\ref{app0n}) is about two orders in magnitude
larger when compared to the maximal value $\varepsilon$, which is Q/2. 
Clearly, in the case of $0\nu\beta\beta$-decay the convergence of the Taylor
expansion of the decay rate is fast and the standard approach given by
the leading term in the Taylor expansion is well justified.

\subsection{Energy distributions of emitted electrons}

The NEMO3 experiment, which ran for seven years before it stopped taking data in 2010,
measured the $2\nu\beta\beta$-decay of $^{100}$Mo with very high statistics of about 
1 million events \cite{expMo}. Due to high statistics of about tens of thousands of events
the currently running EXO \cite{expXea}, KamlandZEN \cite{expXeb} ($^{136}$Xe) and GERDA ($^{76}$Ge)
\cite{expGe} experiments allow precise determination of the $2\nu\beta\beta$-decay energy
distributions as well. A similar statistics is expected to be achieved also by the CUORE
($^{130}$Te) experiment, which has started taking data recently.
New perspectives for analysis of $2\nu\beta\beta$-decay differential characteristics
will be opened by next generation of the double-beta decay experiments like SuperNEMO,
nEXO, Legend, which will contain significantly larger  amount of double beta decay
radioactive source \cite{ROP12,gulpov}. 

\begin{figure}[!t]
  \includegraphics[width=1.05\columnwidth]{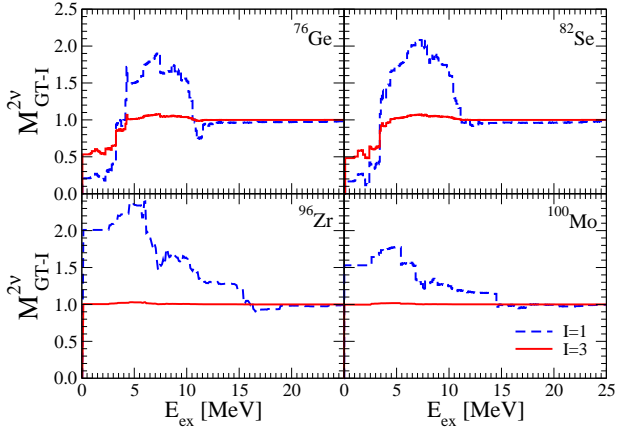}
  \caption{ (Color online) The same as Fig.\ref{fig.runb}
    for the $2\nu\beta\beta$-decay of $^{76}$Ge, $^{82}$Se, $^{96}$Zr and $^{100}$Mo.}
\label{fig.runa}
\end{figure}

By considering the leading first and second order terms in the Taylor expansion
for the single and summed electron differential decay rate normalized
to the  full decay rate we get
\begin{eqnarray}\label{dde}
\frac{1}{\Gamma^{2\nu}} \frac{ d \Gamma^{2\nu}}{d T_{e}} &\simeq&
\frac{1}{\Gamma^{2\nu}} \left(
\frac{ d \Gamma^{2\nu}_0}{d T_{e}} + \frac{ d \Gamma^{2\nu}_2}{d T_{e}} 
\right)\\\label{ddee}
&=& \frac{1}{(G^{2\nu}_0  ~+~ \xi^{2\nu}_{31}~ G^{2\nu}_2)} \left(
\frac{ d G_0}{d T_{e}} ~+~ \xi^{2\nu}_{31}~ \frac{ d G_2}{d T_{e}} 
\right),\nonumber\\
  \frac{1}{\Gamma^{2\nu}} \frac{ d \Gamma^{2\nu}}{d T_{ee}} &\simeq&
\frac{1}{\Gamma^{2\nu}} \left(
\frac{ d \Gamma^{2\nu}_0}{d T_{ee}} + \frac{ d \Gamma^{2\nu}_2}{d T_{ee}} 
\right)\\
&=& \frac{1}{(G^{2\nu}_0  ~+~ \xi^{2\nu}_{31}~ G^{2\nu}_2)} \left(
\frac{ d G_0}{d T_{ee}} ~+~ \xi^{2\nu}_{31}~ \frac{ d G_2}{d T_{ee}} 
\right),
\nonumber
\end{eqnarray}
where
\begin{eqnarray}
  \frac{d G_{N}^{2\nu}}{dT_{e_1}} &=& 
\frac{{c}_{2\nu}}{m_e^{11}}~F_0(Z_f,E_{e_1}) p_{e_1} E_{e_1}\nonumber\\
&&\int_{0}^{Q-T_{e_1}} F_0(Z_f,E_{e_2}) p_{e_2} E_{e_2} I_N(T_{e_1},T_{e_2}) dT_{e_2},
\nonumber\\
  \frac{d G_{N}^{2\nu}}{dT_{ee}} &=& 
\frac{{c}_{2\nu}}{m_e^{11}}~\frac{T_{ee}}{Q} 
\int_{0}^{Q} F_0(Z_f,E_{e_1}) p_{e_1} E_{e_1}\nonumber\\
&&\times F_0(Z_f,E_{e_2}) p_{e_2} E_{e_2} I_N(T_{e_1},T_{e_2}) dV,
\end{eqnarray}
(N=0, 2) with
\begin{eqnarray}
I_N(T_{e_1},T_{e_2}) = \int_{0}^{Q-T_{e_1}-T_{e_2}} E_{\nu_1}^2 E_{\nu_2}^2 {\cal A}^{2\nu}_N dE_{\nu_1}.
\end{eqnarray}
and 
\begin{eqnarray}
T_{ee} = T_{e_1} + T_{e_2},~~~
V = Q \frac{T_{e_2}}{T_{e_1} + T_{e_2}}.
\end{eqnarray}
Here, $E_{\nu_2} = E_i - E_f - E_{e_1}  - E_{e_1}  - E_{\nu_1}$ is determined by 
the energy conservation.
$T_{ee}$ is a sum of kinetic energies of both electrons ($T_{e_1}$ and $T_{e_2}$)
and
$T_e$ represents kinetic energy of any of two emitted electrons.

\begin{figure}[!t]
  \includegraphics[width=1.0\columnwidth]{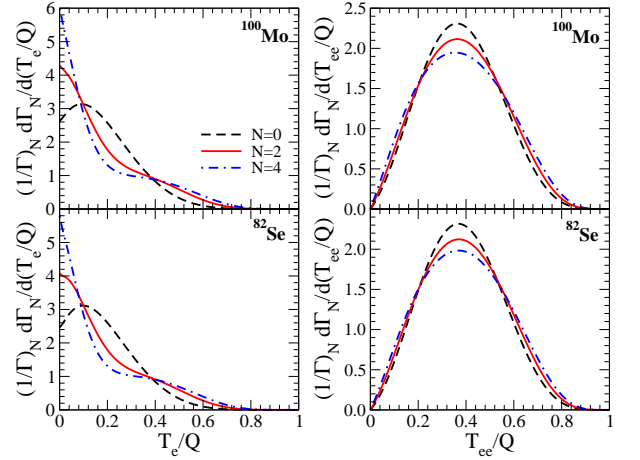}
  \caption{(Color online) The partial differential decay rates $(1/\Gamma_0)~d\Gamma_0/dT_e$,
    $(1/\Gamma_2)~d\Gamma_2/dT_e$ and $(1/\Gamma_4)~d\Gamma_4/dT_e$ normalized
    to corresponding partial decay rate
    vs. kinetic energy of a single electron $T_e$ (in units of Q-value) 
    (left panels) 
    and the partial differential decay rates $(1/\Gamma_0)~d\Gamma_0/dT_{ee}$,
    $(1/\Gamma_2)~d\Gamma_2/dT_{ee}$ and $(1/\Gamma_4)~d\Gamma_4/dT_{ee}$  normalized
    to corresponding partial decay rate 
    vs. the sum of kinetic energies of emitted electrons $T_{ee}$
    (in units of Q-value) (right panels)
    for the $2\nu\beta\beta$-decay of $^{82}$Se  and $^{100}$Mo to ground state of
    final nucleus. The energy distributions are normalized to unity to see
    the differences in shape among them.
  }
\label{fig.shape}
\end{figure}

The single and summed electron differential decay rates 
normalized to the full width in Eqs. (\ref{dde}) and (\ref{ddee})
contain one unknown parameter, namely the ratio $\xi^{2\nu}_{31}$.
We note that  partial contributions to the full differential
decay rate in Eq. (\ref{dde}) (Eq. (\ref{ddee})) exhibit different
behavior as function of $T_e$ ($T_{ee}$). This fact is displayed
in Fig. \ref{fig.shape}, where  single and summed electron
partial differential decay rates normalized to the partial width
(i.e., all energy distributions are normalized to unity and do not depend
on any NME)
are presented for the $2\nu\beta\beta$-decay of $^{82}$Se and $^{100}$Mo.
The difference in
distributions corresponding to the leading and first order terms
in Taylor expansion is apparent especially in the case of single
electron energy distribution. Due to this phenomenon there is
a possibility to deduce ratio $\xi^{2\nu}_{31}$ from the measured
energy distributions.

\begin{figure}[!t]
  \includegraphics[width=1.00\columnwidth]{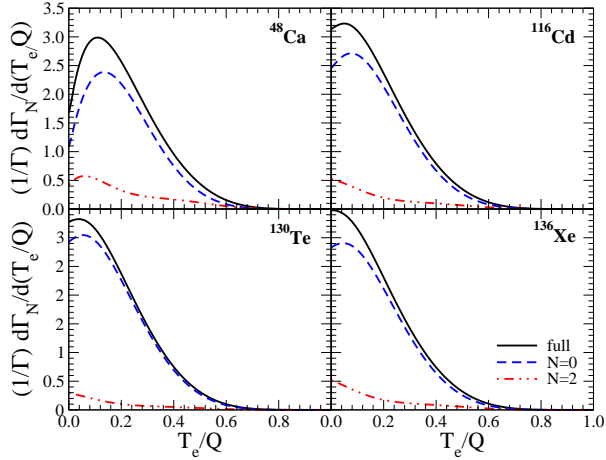}
  \caption{(Color online) The full differential decay rate $(1/\Gamma)~d\Gamma/dT_e$ and partial
    differential decay rates $(1/\Gamma)~d\Gamma_0/dT_e$ and 
    $(1/\Gamma)~d\Gamma_2/dT_e$ normalized to the full decay rate
    vs. the kinetic energy of a single electron $T_e$ (in units of Q-value)  
    for the $2\nu\beta\beta$-decay of $^{48}$Ca, $^{116}$Cd, $^{130}$Te
    and $^{136}$Xe. 
  }
\label{fig.Scacdtexe}
\end{figure}

\begin{figure}[!t]
  \includegraphics[width=1.00\columnwidth]{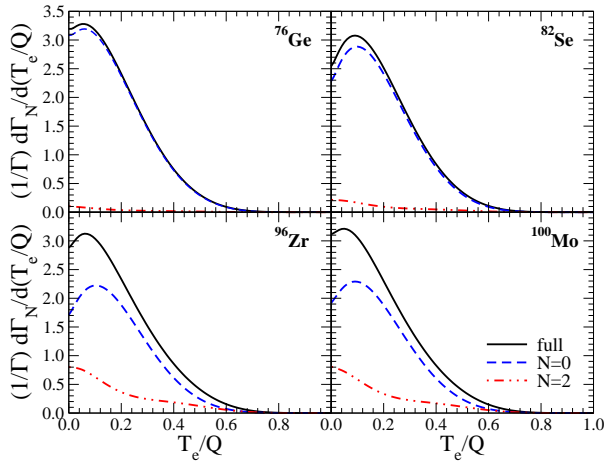}
  \caption{ (Color online) The same as Fig.\ref{fig.Scacdtexe}
    for the $2\nu\beta\beta$-decay of $^{76}$Ge, $^{82}$Se, $^{96}$Zr and $^{100}$Mo.
  }
\label{fig.Sgesezrmo}
\end{figure}

For pn-QRPA value of the parameter $\xi^{2\nu}_{31}$ (see Table \ref{tab:nme})
the full differential decay rate $(1/\Gamma^{2\nu})~d\Gamma^{2\nu}/dT_e$ and partial
differential decay rates $(1/\Gamma^{2\nu})~d\Gamma^{2\nu}_0/dT_e$, $(1/\Gamma^{2\nu})~d\Gamma^{2\nu}_2/dT_e$
normalized to the full decay rate  are presented
as function of the  kinetic energy of a single electron $T_e$
(sum of kinetic energy of both electrons $T_{ee}$
for the eight $2\nu\beta\beta$-decay isotopes
in Figs. \ref{fig.Scacdtexe} and \ref{fig.Sgesezrmo} 
(\ref{fig.Tcacdtexe} and \ref{fig.Tgesezrmo}).
We see that the largest contribution from the additional term due to Taylor expansion to
the full differential decay rate is found by the $2\nu\beta\beta$-decay of
$^{100}$Mo, $^{96}$Zr, $^{48}$Ca, $^{116}$Cd and $^{136}$Xe. These isotopes are good
candidates to measure $\xi^{2\nu}_{31}$ in double beta decay experiments.

\begin{figure}[!t]
  \includegraphics[width=1.00\columnwidth]{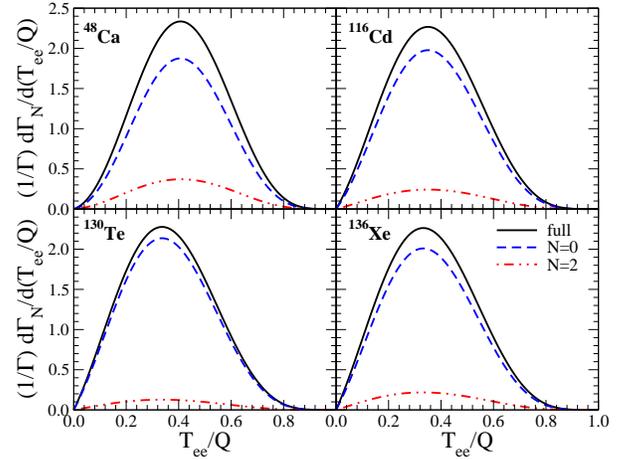}
  \caption{(Color online) The full differential decay rate $(1/\Gamma)~d\Gamma/dT_e$ and partial
    differential decay rates $(1/\Gamma)~d\Gamma_0/dT_e$ and
    $(1/\Gamma)~d\Gamma_2/dT_e$ normalized to the full decay rate
    vs. the sum of kinetic energies of emitted electrons $T_{ee}$ (in units of Q-value) 
    for the $2\nu\beta\beta$-decay of $^{48}$Ca, $^{116}$Cd, $^{130}$Te
    and $^{136}$Xe.
  }
\label{fig.Tcacdtexe}
\end{figure}

\begin{figure}[!t]
  \includegraphics[width=1.00\columnwidth]{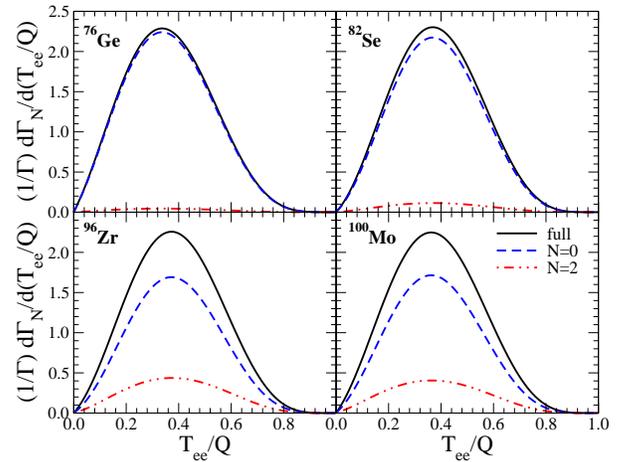}
  \caption{(Color online) The same as Fig.\ref{fig.Tcacdtexe}
    for the $2\nu\beta\beta$-decay of $^{76}$Ge, $^{82}$Se, $^{96}$Zr and $^{100}$Mo.
  }
\label{fig.Tgesezrmo}
\end{figure}

By assuming  $\xi^{2\nu}_{13} =$ 0.0, 0.4 and 0.8. the  single electron energy  distribution and
summed electron  energy  spectrum normalized to the full decay rate
for $2\nu\beta\beta$-decay of $^{82}$Se and $^{100}$Mo are presented in Fig. \ref{fig.tough}.
We see that corresponding curves are close to each other and that high statistics
of the $2\nu\beta\beta$-decay experiment is needed to deduce information about the ratio
of nuclear matrix elements $\xi^{2\nu}_{13}$ from the data. The study performed within
the NEMO3 experiment \cite{SSDexp} in respect the SSD versus HSD
hypthesis \cite{domin1,domin2} has shown  that it is feasable. It might be that
the hight statistics achieved by the GERDA \cite{expGe}, CUORE \cite{expTe}, 
EXO ($^{136}$Xe) and KamlandZEN ($^{136}Xe$) experiments is sufficient to
conclude about the value of $\xi^{2\nu}_{13}$ for the measured $2\nu\beta\beta$-decay
transition.

\begin{figure}[!t]
  \includegraphics[width=1.00\columnwidth]{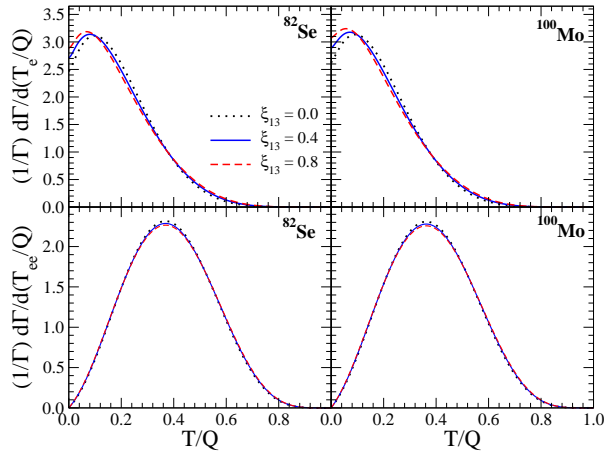}
  \caption{(Color online) Differential decay rates $(1/\Gamma)~d\Gamma/d(T_e/Q)$
(upper panels) and $(1/\Gamma)~d\Gamma/d(T_{ee}/Q)$ (lower panels) 
normalized to full decay rate $\Gamma$  vs. kinetic energy of a single electron $T=T_e$
and the sum of kinetic energies of emitted electrons $T=T_{ee}$ (in units of Q-value),
respectively. Results are presented for the $2\nu\beta\beta$-decay
of $^{82}$Se (left panels) and $^{100}$Mo (right panels) by assuming
$\xi^{2\nu}_{13} =$ 0.0, 0.40 and 0.8. }
\label{fig.tough}
\end{figure}

For some of future double-beta decay experiments the $2\nu\beta\beta$-decay is considered as
important background for the signal of the $0\nu\beta\beta$-decay, e.g., in the case of the
SuperNEMO experiment. In Fig. \ref{fig.ends} the endpoint of the spectrum of the
differential decay rate normalized to the full  decay rate 
$(1/\Gamma)~d\Gamma/dT$ as function of the sum of kinetic energy of emitted electrons
$T = (E_{e_1} + E_{e_2} - 2m_e)$  is presented for the $2\nu\beta\beta$-decay of $^{82}$Se
and $^{100}$Mo. The results were obtained with the common and improved theoretical
expressions for the $2\nu\beta\beta$-decay rate. We see that by considering
revised formula the number of the $2\nu\beta\beta$-decay events close to the end
of spectra is slightly suppressed in comparison with previous expectations, 
what is apparent especially in the case of the $2\nu\beta\beta$-decay of $^{100}$Mo.

\begin{figure}[!t]
  \includegraphics[width=1.00\columnwidth]{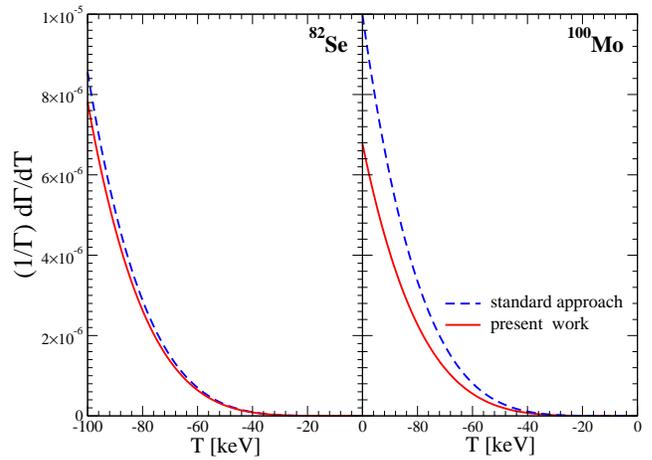}
  \caption{ (Color online) The endpoint of the spectrum of the
    differential decay rate normalized to the full  decay rate
    $(1/\Gamma)~d\Gamma/dT$ vs. the sum of kinetic energy of emitted electrons
    $T = (E_{e_1} + E_{e_2} - 2m_e)$  
    for the $2\nu\beta\beta$-decay of $^{82}$Se  and $^{100}$Mo.
    The calculation with the standard (leading term in Taylor expansion)
    and improved (present work) theoretical description of the $2\nu\beta\beta$-decay rate.
    The considered ratios $\xi^{2\nu}_{31}$ and $\xi^{2\nu}_{51}$ are
    those calculated within QRPA with isospin restoration
    (see Table \ref{tab:nme}).    
  }
\label{fig.ends}
\end{figure}

\subsection{Evaluation of the effective axial-vector coupling constant}

The calculation of $M^{2\nu}_{GT-3}$ can be more reliable as that of $M^{2\nu}_{GT-1}$,
because $M^{2\nu}_{GT-3}$ is saturated
by contributions through the lightest states of the intermediate
nucleus. Thus, we rewrite the $2\nu\beta\beta$-decay rate 
as follows:
\begin{eqnarray}
  \left[T^{2\nu \beta \beta}_{1/2}\right]^{-1} \simeq \left(g^{\rm eff}_A\right)^4
  \left| M^{2\nu}_{GT-3}\right|^2 \frac{1}{\left|\xi^{2\nu}_{31} \right|^2}
  \left(G^{2\nu}_{0}  + \xi^{2\nu}_{31}  G^{2\nu}_{2} \right),
\label{tga}  \nonumber\\
\end{eqnarray}
i.e., without explicit dependence on matrix element $M^{2\nu}_{GT-1}$. 
For sake of simplicity it is assumed that values of  involved nuclear matrix
elements are  real. From Eq. (\ref{tga}) it follows that 
if $\xi^{2\nu}_{31}$ is deduced from the measured $2\nu\beta\beta$-decay
energy distribution and $M^{2\nu}_{GT-3}$ is reliably calculated 
by nuclear structure theory, the value of the effective axial-vector coupling constant
$g_A^{\rm eff}$ can be determined from the measured $2\nu\beta\beta$-decay
half-life. 

\begin{figure}[!t]
  \includegraphics[width=1.05\columnwidth]{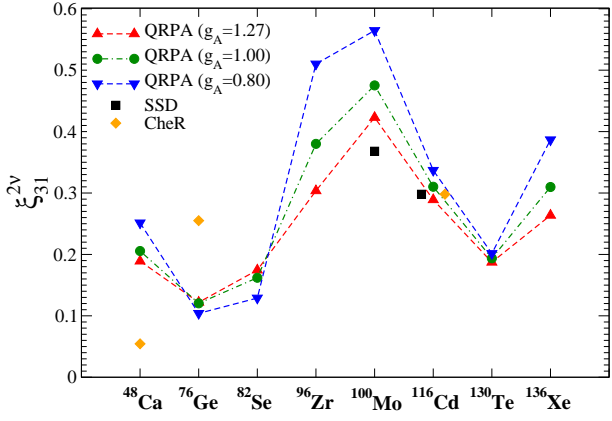}
  \caption{(Color online) The ratio $\xi^{2\nu}_{31}$ of nuclear matrix elements
$M^{2\nu}_{GT-3}$ and $M^{2\nu}_{GT-1}$ calculated within the pn-QRPA 
    with partial restoration of isospin symmetry \cite{vadimp} by
    assuming $g_A^{\rm eff}$=0.80, 1.00 1nd 1.269, the single state
    dominance hypothesis (SSD) \cite{domin1,domin2} and by using the Gamow-Teller strengths
    measured in charge-exchange reactions (CheR) \cite{frek1,frek2,frek3,frek4}
    under the assumption of equal phases of all contributions to the matrix element.
  }
\label{fig.xi}
\end{figure}

Let discuss the value of $\xi^{2\nu}_{31}$ within different approaches
before it will be measured by the double-beta decay experiment. 
Within the SSD hypothesis \cite{domin1,domin2,abad} it is supposed
that the $2\nu\beta\beta$-decay NME is governed by the two virtual transitions:
the first one going from the initial $0^+$ ground state to the $1^+$ ground state
of the intermediate nucleus and second one from this $1^+$ state to the final
$0^+$ ground state.  Within this assumption we obtain
\begin{eqnarray}
  (g_A^{\rm eff})^2 M^{2\nu}_{GT-k} &\simeq& m^k_e
  \frac{(g_A^{\rm eff})^2 M_1}{\left(E_1 - \frac{(E_i-E_f)}{2}\right)^k} \\
  &=& \frac{3 D}{\sqrt{ft_\beta ~ft_{EC}}} \frac{m_e^k}{\left(E_1 - \frac{(E_i-E_f)}{2}\right)^k}
  \nonumber
\end{eqnarray}
with k=1 and 3. Here, $D=(3 \pi^3 \ln{(2)})/(G^2_\beta m_e^5)$ is the
beta decay constant. The main advantage of the SSD approach is that the product 
$(g_A^{\rm eff})^2M_1$ can be evaluated from the measured $log ft$ values
associated with the electron capture and single $\beta$-decay of the ground state
of intermediate nucleus with $J^\pi = 1^+$. There are three double beta systems
with A=100, 116 and 128, which allow it. The corresponding SSD predictions
for $(g_A^{\rm eff})^2 M^{2\nu}_{GT-k}$ (k=1 and 3) and $\xi^{2\nu}_{31}$
are listed in Table \ref{tab.phen}.

The Gamow-Teller strengths to excited states of intermediate nucleus
from initial and final ground states entering the double beta decay
transition are measured with help of charge-exchange
reactions (ChER) \cite{frek1,frek2,frek3,frek4}, i.e.,
via strong interaction due to spin-isospin Majorana force.
For $^{48}$Ca, $^{76}$Ge and $^{116}$Cd the calculated matrix elements
$M^{2\nu}_{GT-1}$, $M^{2\nu}_{GT-3}$ and $\xi^{2\nu}_{31}$ under the assumption of
equal phases for its each individual contribution are presented in 
Table \ref{tab.phen}. The CheR allow to measure with a reasonable
resolution of about tens of keV the Gamow-Teller strengths 
only up to about 5 MeV, i.e., below the region of the Gamow-Teller resonance,
what might be considered as drawback. We note  that some questions arise
also about the normalization of the Gamow-Teller strengths by the experiment. 

The pnQRPA, SSD and CheR predictions for parameter $\xi^{2\nu}_{31}$ for various
isotopes are displayed in Fig. \ref{fig.xi}. We see that a best agreement among different
results occurs by $^{116}$Cd. In the case of $^{48}$Ca and $^{76}$Ge there is a
significant difference between the pn QRPA and CheR results. We note that within
the HSD hypothesis \cite{domin1,domin2} the value of  $\xi^{2\nu}_{31}$ is equal to zero. 

By considering  the SSD values for $\xi^{2\nu}_{31}$ 
(see Table \ref{tab.phen}) we obtain
\begin{eqnarray}\label{gaequation}
  g^{\rm eff}_A (^{100}{\rm Mo}) = \frac{0.251}{\sqrt{M^{2\nu}_{GT-3}}}, ~~~~
  g^{\rm eff}_A (^{116}{\rm Cd}) = \frac{0.214}{\sqrt{M^{2\nu}_{GT-3}}}.\nonumber\\
\end{eqnarray}  
The corresponding curves are plotted in Fig. \ref{fig.mgt3}. It is apparent that 
if the value of $M^{2\nu}_{GT-3}$ would be calculated reliably, e.g. within
the interacting shell model, which is known to describe very well the lowest
excited states of parent
and daughter nucleus participating in double-beta decay process, one could conclude
about the value of the effective axial-vector coupling constant $g_A^{\rm eff}$
for a given nuclear system. However, we note that the correct value of $g_A^{\rm eff}$
can be determined only if $\xi^{2\nu}_{31}$ deduced from the measured
$2\nu\beta\beta$-decay energy distribution is considered. In that case 
the constant on the r.h.s of Eq. (\ref{gaequation}) might be different.

\begin{widetext}

\begin{table*}[htb]  
  \begin{center}  
    \caption{The nuclear matrix elements $M^{2\nu}_{\rm GT-1}$ and $M^{2\nu}_{\rm GT-3}$
      calculated from measured $GT^\pm$ strengths in charge exchange reaction (ChER) under the
      assumption of a equal phases for its each individual contribution \cite{frek1,frek2,frek3,frek4}
      and their
      product with squared effective axial-vector coupling constant $g^{\rm eff}_A$,
      which is determined within the Single State Dominance Hypothesis (SSD hypothesis) \cite{domin1,domin2}.
    }  
\label{tab.phen}  
\renewcommand{\arraystretch}{1.2}  
\begin{tabular}{lcccccccccccc}\hline\hline
 & & \multicolumn{5}{c}{SSD} & & \multicolumn{5}{c}{ChER} \\ 
\cline{3-7} \cline{9-13}                       
Nucl. & & $(g^{\rm eff}_{\rm A})^2 M^{2\nu}_{\rm GT-1}$ & $(g^{\rm eff}_{\rm A})^2 M^{2\nu}_{\rm GT-3}$ &
            $(g^{\rm eff}_{\rm A})^2 M^{2\nu}_{\rm GT-5}$ & $\xi^{2\nu}_{31}$ & $\xi^{2\nu}_{51}$  & &
 $M^{2\nu}_{\rm GT-1}$ & $M^{2\nu}_{\rm GT-3}$ &  $M^{2\nu}_{\rm GT-5}$ & $\xi^{2\nu}_{31}$ & $\xi^{2\nu}_{51}$  \\ \hline
$ ^{48}$Ca  & &       -        &     -         &       -       &      -         &       -      & &
                $4.25\times 10^{-2}$ & $2.31\times 10^{-3}$ & $1.26\times 10^{-4}$ & $0.054$ & $0.003$ \\
$ ^{76}$Ge  & &       -        &     -         &       -       &      -         &       -      & &
                $8.61\times 10^{-2}$ & $2.20\times 10^{-2}$ & $5.61\times 10^{-3}$ & $0.255$ & $0.065$ \\
$^{100}$Mo  & & $1.71\times 10^{-1}$ & $6.29\times 10^{-2}$ & $2.31\times 10^{-2}$ & $0.368$ & $0.135$ & &
                    -        &     -         &       -       &      -         &   -  \\
$^{116}$Cd  & & $1.53\times 10^{-1}$ & $4.57\times 10^{-2}$ & $1.36\times 10^{-2}$ & $0.298$ & $0.089$ & &
               $5.88\times 10^{-2}$ & $1.75\times 10^{-2}$ & $5.22\times 10^{-3}$ & $0.298$ & $0.089$ \\
$^{128}$Te  & &  $1.60\times 10^{-2}$ & $5.87\times 10^{-3}$ & $2.16\times 10^{-3}$ & $0.367$ & $0.135$ & &
                    -        &     -         &       -       &      -         &   -  \\
\hline\hline
\end{tabular}\\
  \end{center}  
\end{table*}  

\end{widetext}      

\begin{figure}[!t]
  \includegraphics[width=1.10\columnwidth]{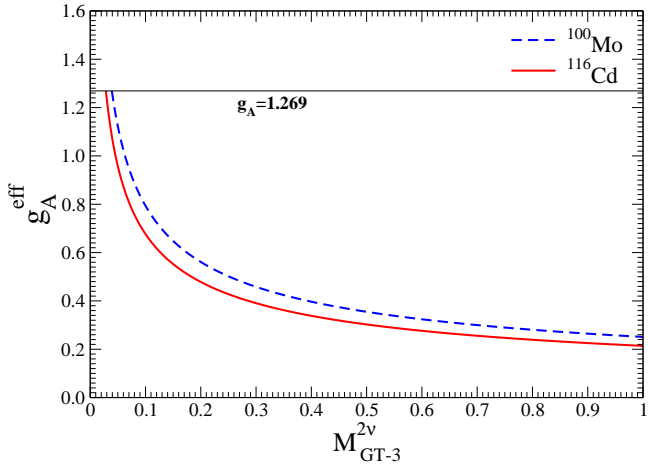}
  \caption{(Color online) The effective axial-vector coupling constant $g^{\rm eff}_A$
    as function of the matrix element $M^{2\nu}_{GT-3}$ for
    $2\nu\beta\beta$-decay of $^{100}$Mo  and $^{116}$Cd.
    The SSD values are assumed for $\xi^{2\nu}_{31}$ (see Table \ref{tab.phen}).
  }
\label{fig.mgt3}
\end{figure}

%
\section{Summary and conclusions}
%

In summary, improved formulae for the $2\nu\beta\beta$- and
$0\nu\beta\beta$-decay half-lives are presented by taking advantage
of the Taylor expansion over the parameters containing the lepton
energies of energy denominators. The additional terms
due to Taylor expansion in the decay rate have been found significant
in the case of the $2\nu\beta\beta$-decay and practically of no importance
in the case of the $0\nu\beta\beta$-decay.

Up to first order in the Taylor expansion the $2\nu\beta\beta$-decay
rate includes two nuclear matrix elements $M^{2\nu}_{GT-1}$ and $M^{2\nu}_{GT-3}$
with energy denominator in the first and third power, respectively.
It was shown that the ratio of these matrix elements
$\xi^{2\nu}_{31}= M^{2\nu}_{GT-3}/M^{2\nu}_{GT-1}$ might be determined experimentally from
the shape of the single and sum electron energy distributions, if the statistics
of a considered double beta decay experiment allows it. A study of the 
SSD and HSD hypotheses in the case of the  $2\nu\beta\beta$-decay of $^{100}$Mo
by the NEMO3 experiment has manifested that it is feasible \cite{SSDexp}. 

A measured value
of $\xi^{2\nu}_{31}$ is expected to be an important information about
virtual transitions through the states of intermediate nucleus. 
The calculation of running sum of $M^{2\nu}_{GT-1}$ and $M^{2\nu}_{GT-3}$
performed within the pn-QRPA with partial restoration of isospin
symmetry showed that $M^{2\nu}_{GT-3}$ is determined 
by contributions through the low-lying states of the intermediate
nucleus unlike $M^{2\nu}_{GT-1}$, which is affected significantly also
by contributions through transitions over intermediate nucleus from
the region of the Gamow-Teller resonance.

Further, the $2\nu\beta\beta$-decay
rate was expressed with $M^{2\nu}_{GT-3}$ and $\xi^{2\nu}_{31}$, i.e.
without the explicit dependence on the commonly studied nuclear matrix
element$M^{2\nu}_{GT-1}$. It was suggested that one can get information
about the axial-vector coupling constant in nuclear medium $g_A^{\rm eff}$ 
once $\xi^{2\nu}_{31}$ is deduced from the measured electron energy distribution
and $M^{2\nu}_{GT-3}$ is calculated reliably, e.g. within the ISM. 

It goes without saying that improved formula for the $2\nu\beta\beta$-decay
half-life will play an important role in accurate analysis of the
Majoron mode of the $0\nu\beta\beta$-decay and study of Lorentz invariance
violation, bosonic admixture of neutrinos and other effects. 

\begin{acknowledgments}
This work is supported by the VEGA Grant Agency of the Slovak Republic under
Contract No. 1/0922/16, by Slovak Research and Development Agency under Contract
No. APVV-14-0524, RFBR Grant No. 16-02-01104, Underground laboratory
LSM - Czech participation to European-level research
infrastructure CZ.02.1.01/0.0/0.0/16 013/0001733.
\end{acknowledgments}

%

%



\begin{thebibliography}{99}

\bibitem{hax84} W.C. Haxton and G.J. Stephenson, Jr., Prog. Part. Nucl. Phys. {\bf 12}, 409 (1984).

\bibitem{doi85} M. Doi, T. Kotani, E. Takasugi, Prog. Theor. Phys. Suppl. {\bf 83}, 1 (1985).
  
\bibitem{ROP12} J.D. Vergados, H. Ejiri, and F. \v Simkovic, Rep. Prog. Phys. {\bf 75}, 106301 (2012).   

\bibitem{barabash} A.S. Barabash,  Nucl. Phys. A {\bf 935}, 52 (2015).

\bibitem{dussu4} D. \v{S}tefanik, F. \v{S}imkovic, and A. Faessler, Phys. Rev. C {\bf 91}, 064311 (2015).
  
\bibitem{Rod03a} V.A.~Rodin, A.~Faessler, F.~\v Simkovic and P.~Vogel,
   Phys. Rev. C {\bf 68}, 044302 (2003).  
  
\bibitem{Rod06} V.A.~Rodin, A.~Faessler, F.~\v Simkovic and P.~Vogel,
Nucl. Phys.{\bf A766}, 107 (2006) and erratum, 
Nucl. Phys. {\bf A793}, 213 (2007).

\bibitem{LSSM1}   E. Caurier, F. Nowacki, A. Poves, J. Retamosa,
            Phys. Rev. Lett. {\bf 77}, 1954 (1996).

\bibitem{kmuto} K. Muto, E. Bender, and H.V. Klapdor-Kleingrothaus,
   Z. Phys. A {\bf 339}, 435 (1991).            

\bibitem{majoron} The NEMO-3 Collab., R. Arnold, et al., Nucl. Phys. A {\bf 765}, 483 (2006). 

\bibitem{bosneu} A.S. Barabash, A.D. Dolgov, R. Dvornick\'{y}, and F. \v{S}imkovic,
  Nucl. Phys. B {\bf 783}, 90 (2007). 

\bibitem{lorentz} The EXO-200 Collab., J.B. Albert, et al.,
  Phys. Rev. D {\bf 93}, 072001 (2016).
  
\bibitem{domin1} F.\v{S}imkovic, P. Domin, and S.V. Semenov,
  J. Phys. G {\bf 27}, 2233 (2001). 

\bibitem{domin2} P. Domin, S. Kovalenko, F.\v{S}imkovic, and S.V. Semenov, 
  Nucl. Phys. A {\bf 753}, 337 (2005).

\bibitem{expGe}
  The GERDA Collab., M. Agostini, et al., Nature  544, {\bf 47}  (2017).

\bibitem{expMo} 
  The NEMO-3 Collab., R. Arnold, et al., Phys. Rev. D {\bf 92}, 072011 (2015). 

\bibitem{expTe} 
The CUORE Collab., K. Alfonso, et al., Phys. Rev. Lett. {\bf 115}, 102502 (2015).

\bibitem{expXea}
The EXO Collab.,  N. Ackerman, et al., Phys. Rev. Lett. {\bf 107}, 212501 (2011).
    
\bibitem{expXeb} 
The KamLAND-Zen Collab., A. Gando, et al., Phys. Rev. C {\bf 85}, 045504 (2012).

\bibitem{ibmcl}
  J. Barea, J. Kotila, and F. Iachello, Phys.Rev. C {\bf 91}, 034304 (2015).

\bibitem{abad} 
 J. Abad, A. Morales, R. Nunez-Lagos and A. Pacheco, 
 Ann. Fis. A {\bf 80}, 9 (1984).

\bibitem{javierga} J. Menendez, D. Gazit, and A. Schwenk,
  Phys. Rev. Lett. {\bf 107}, 062501 (2011). 
  
\bibitem{engelga} J. Engel, F. \v{S}imkovic, and P. Vogel, Phys.Rev. C {\bf 89}, 064308 (2014).

 
\bibitem{iachello} J. Kotila and F. Iachello, Phys. Rev. C {\bf 85}, 034316 (2012).

\bibitem{radial} F. Salvat, J.M. Fernandez-Varea, W. Williamson Jr.,
     Comput. Phys. Commun. {\bf 90}, 151 (1995).  

\bibitem{vadimp} F. \v Simkovic, V. Rodin, A. Faessler, and P. Vogel, Phys. Rev. C {\bf 87}, 045501 (2013). 

\bibitem{gulpov} A. Giuliani and A. Poves, Adv. High Energy Phys. {\bf 2012}, 857016 (2012). 
  
\bibitem{frek1} 
 S. Rakers, et al., Phys. Rev .C {\bf 70}, 054302 (2004).

\bibitem{frek2} 
 E.W. Grewe, et al., Phys. Rev. C {\bf 78}, 044301 (2008).

\bibitem{frek3} 
 H. Dohmann, et al., Phys. Rev. C {\bf 78}, 041602(R) (2008).

\bibitem{frek4} 
 S. Rakers, et al., Phys. Rev. C {\bf 71}, 054313 (2005).

\bibitem{SSDexp} 
  The NEMO-3 Collab., R. Arnold, et al., JETP Letters {\bf 80}, 377 (2004).


\end{thebibliography}
\end{document}